\documentclass[aps,superscriptaddress,preprint]{revtex4-1}
\usepackage{latexsym}
\usepackage{amssymb}
\usepackage{graphicx,bm}
\usepackage[utf8]{inputenc}
\newcommand{\ba}{\begin{eqnarray}} \newcommand{\ea}{\end{eqnarray}}
\newcommand{\be}{\begin{equation}} \newcommand{\ee}{\end{equation}}

\usepackage{amsmath,amssymb,amsthm,graphicx,bm,color}
\usepackage{slashed}
\usepackage{amsmath,amssymb,amsthm,graphicx,bm,color}

\renewcommand{\Delta}{\varDelta} 
\renewcommand{\Gamma}{\varGamma} 
\renewcommand{\Omega}{\varOmega} 
\renewcommand{\Phi}{\varPhi} 
\renewcommand{\Psi}{\varPsi} 
\renewcommand{\Sigma}{\varSigma} 
\renewcommand{\Theta}{\varTheta} 
\renewcommand{\epsilon}{\varepsilon}

\begin{document}

\title{Flipped Quartification and a composite $b$-quark}

\author{James B. Dent}\email{jbdent@shsu.edu}

\affiliation{Department of Physics, Sam Houston State University, Huntsville, TX 77341, USA}

\author{Thomas W. Kephart}\email{tom.kephart@gmail.com}

\affiliation{Department of Physics and Astronomy, Vanderbilt
University, Nashville, TN 37235} 

\author{Heinrich P\"as}\email{heinrich.paes@uni-dortmund.de}

\affiliation{ Institut f\"ur Physik, Technischen Universit\"at Dortmund,
D-44221 Dortmund, Germany} 

\author{Thomas J. Weiler}\email{tom.weiler@vanderbilt.edu}
\affiliation{Department of Physics and Astronomy, Vanderbilt
University, Nashville, TN 37235}

\date{\today}

\begin{abstract}
 An alternative ``flipped" version of the  quartification model is obtained by rearrangement of the particle assignments. The model has two standard (trinification) families and one flipped quartification family. An interesting phenomenological implication is that the model allows for a composite b-quark.
 \end{abstract}


\maketitle

\newpage
\section{Introduction}

The model presented here, which we call ``Flipped Quartification,'' is an
extension of the
standard model (SM) that singles out the $b$ quark as different from
all the rest of the SM fermions in that just above the electro-weak
(EW) scale the EW singlet $b_R$ can be in
a nontrivial irreducible representation (irrep) of a new gauge group
$SU(2)_{\ell}$ while all
the other fermions are in $SU(2)_{\ell}$ singlets. This can happen
within the model in two ways:
(i) the $SU(2)_{\ell}$ symmetry breaks just above the EW scale where
now the $b_R$
falls into its usual SM irrep, but with slightly different
phenomenology due to nearby $SU(2)_{\ell}$
effects that the other fermions do not have. This is a fairly
conventional scheme for introducing
new physics into the SM.  More interesting is (ii) where
$SU(2)_{\ell}$ becomes confining
just above the EW scale. This is possible for a range parameters
chosen at the unification
scale of quartification where all gauge symmetries are restored.\\

Trinification models,
\cite{deRujula:1984,Babu:1985gi,He:1986cs,Nishimura:1988fp,Carlson:1992ew,Lazarides:1993uw,Lazarides:1994px,Willenbrock:2003ca,Choi:2003ag,Kim:2004pe,Carone:2004rp,Carone:2005ha,Demaria:2005gka,Carone:2005rk} with gauge group $SU(3)_L \times SU(3)_C \times
SU(3)_R$, 
and 
quartification  models
\cite{Joshi:1991yn,Babu:2003nw,Chen:2004jza,Demaria:2005gka,Demaria:2006uu,Demaria:2006bd,Babu:2007fx}, where 
the gauge group is extended to 
$SU(3)_l \times SU(3)_L \times SU(3)_C \times SU(3)_R$,
are both in a class of models where the fermions are in bifundamental representations.  Here we will concentrate on the phenomenology of a new class of quartification models obtained by "flipping" the $SU(3)_l $ and $SU(3)_R$ groups.

All quartification  models contain a leptonic color sector 
to realize a manifest quark-lepton symmetry 
\cite{Foot:1990dw,Foot:1991fk,Foot:2006ie}
and must contain at 
least three  families to be phenomenologically viable, plus they 
contain the new fermions needed to symmetrize the quark and lepton particle 
content at high energies.   
Instead of  fully quartified models, where all families are quartification families are given by
\be{}
3 [(3\bar{3}11)+ (13\bar{3}1) +  (113\bar{3}) +  (\bar{3}113)],
\ee
we will consider only hybrid models 
\be{}
n [(13\bar{3}1) + (113\bar{3}) + (1\bar{3}13)] +
(3-n)[(3\bar{3}11)+ (13\bar{3}1) +  (113\bar{3}) +  (\bar{3}113)].
\ee
where $n>0$ families are trinification families 
and the the remaining $3-n$ are 
 quartification families.
In particular, we concentrate on the $n=2$ case \cite{Babu:2007fx}.

One can derive three family models with   appropriate scalar content 
to permit gauge symmetry breaking to the standard model and ultimately to 
$SU(3)_C\times U_{EM}(1)$ from orbifolded $AdS\otimes S^5$ (for a review see \cite{Frampton:2007fr}). In \cite{Kephart:2001qu,Kephart:2004qp} two of us 
carried out a global search for  $\Gamma=Z_n $ trinification models with three or more families, and in \cite{Babu:2007fx} quartification  models of this type were derived from a $\Gamma=Z_8 $ orbifolded $AdS\otimes S^5$.  We leave the study of the UV completion of this model for later work.

\section{Flipped $2+1$ quartification model}
Under the original quartification gauge group $SU(3)_{l}\times SU(3)_{L}\times SU(3)_{C} \times SU(3)_{R}$ the representations 
of the two trinification plus one quartification family model (the $2+1$ Q-model of reference \cite{Babu:2007fx}) were given by
\begin{eqnarray}
2[(13\bar{3}1) + (113\bar{3}) + (1\bar{3}13)] + [(3\bar{3}11)+(13\bar{3}1)+(113\bar{3})+(\bar{3}113)]
\end{eqnarray}
We now ``flip'' the $R$ and $l$ designations such that
\begin{eqnarray}
lLCR \rightarrow RLCl.
\end{eqnarray}
We are free to cyclically permute the groups and to reverse their order without changing the physics. Thus
we let
\begin{eqnarray}
RLCl \rightarrow CLRl
\end{eqnarray}
which allows us to write our new $2+1$ flipped quartification model in a form that conforms with the notation of earlier work.
Symmetry breaking can easily be arranged with a single adjoint scalar VEV for each of $SU(3)_L$ and  $SU(3)_l$ and a pair of adjoints for $SU(3)_R$ such that
\begin{eqnarray}
SU(3)_L &&\rightarrow SU(2)_L \times U(1)_A\\
SU(3)_R &&\rightarrow U(1)_B\times U(1)_C\\
SU(3)_l &&\rightarrow SU(2)_l \times U(1)_D
\end{eqnarray}
where the charge operator $A$, $C$ and $D$ are of the form $diag(1,1,-2)$ and $B$ is of the  form $diag(1,-1,0)$. Their weighting in forming weak hypercharge will be provided below.

Under the remaining symmetry group $SU(3)_C\times SU(2)_L\times SU(2)_l\times U(1)_A\times U(1)_B\times U(1)_C \times U(1)_D$ the first two families decomposed as in a standard trinification model, 
\begin{eqnarray}
&&(3 \bar{3} 1 1) \rightarrow (3 2 1)_{-1 0 0 0} + (3 1 1)_{2 0 0 0}\\\nonumber
&&(1 3 \bar{3} 1) \rightarrow (1 2 1)_{1 -1 -1 0} + (1 2 1)_{1 1 -1 0} + (1 2 1)_{1 0 2 0} + (1 1 1)_{-2 -1 -1 0} + (1 1 1)_{-2 1 -1 0} + (1 1 1)_{-2 0 2 0}\\\nonumber
&&(\bar{3} 1  3 1) \rightarrow (\bar{3} 1 1)_{0 110} + (\bar{3} 1 1)_{0 -110}+ (\bar{3} 1 1)_{0 0  -20}
\end{eqnarray}
while the third family representations become
\begin{eqnarray}
&&(3 \bar{3} 1 1) \rightarrow (3 2 1)_{-1 0 0 0} + (3 1 1)_{2 0 0 0}\\\nonumber
&&(1 3 \bar{3} 1) \rightarrow (1 2 1)_{1 -1 -1 0} + (1 2 1)_{1 1 -1 0} + (1 2 1)_{1 0 2 0} + (1 1 1)_{-2 -1 -1 0} + (1 1 1)_{-2 1 -1 0} + (1 1 1)_{-2 0 2 0}\\\nonumber
&&(1 1 3 \bar{3}) \rightarrow (1 1 2)_{0 1 1 -1} + (1 1 2)_{0 -1 1 -1} + (1 1 2)_{0 0 -2 -1} + (1 1 1)_{0 1 1 2} + (1 1 1)_{0 -1 1 2} + (1 1 1)_{0 0 -2 2}\\\nonumber
&&(\bar{3} 1 1 3) \rightarrow (\bar{3} 1 2)_{0 0 0 1} + (\bar{3} 1 1)_{0 0 0 -2} \label{3rdFamily}
\end{eqnarray}
Using the relation
\begin{eqnarray}
Q = T_3 + Y
\end{eqnarray}
where $Q$ is the electric charge, $T_3$ is the third component of isospin, and $Y$ is the hypercharge, we can determine the hypercharge in terms of the $U(1)$ charges (designated by $A,B,C$, and $D$) as
\begin{eqnarray}
Y = -\frac{1}{6}A + \frac{1}{2}B -\frac{1}{6}C + \frac{1}{3}D
\end{eqnarray}

Charged singlets can be used to break $U(1)_A\times U(1)_B\times U(1)_C \times U(1)_D$ to the standard weak hypercharge $U(1)_Y$ resulting in
\begin{eqnarray}
&&(3 \bar{3} 1 1) \rightarrow (3 2 1)_{\frac{1}{6}} + (3 1 1)_{-\frac{1}{3}}\\\nonumber
&&(1 3 \bar{3} 1) \rightarrow  (1 2 1)_{-\frac{1}{2}} + (1 2 1)_{\frac{1}{2}} + (1 2 1)_{\frac{1}{2}} + (1 1 1)_{0} + (1 1 1)_{1} + (1 1 1)_{0}\\\nonumber
&&(\bar{3} 1  3 1) \rightarrow (\bar{3} 1 1)_{\frac{1}{3}} + (\bar{3} 1 1)_{-\frac{2}{3}}+ (\bar{3} 1 1)_{\frac{1}{3}}
\end{eqnarray}
for the first two families, where  as usual, each trinification family contains a SM family
\begin{eqnarray}
&& Q^{1(2)}_L+d(s)_R+u(c)_R+l^{1(2)}_L+e(\mu)_R=(3 2 1)_{\frac{1}{6}} + (\bar{3} 1 1)_{\frac{1}{3}} + (\bar{3} 1 1)_{-\frac{2}{3}} + (1 2 1)_{\frac{1}{2}}+ (1 1 1)_{1}
\end{eqnarray}
plus the following vector-like  states
\begin{eqnarray}
&&+ (\bar{3} 1 1)_{\frac{1}{3}}+ (3 1 1)_{-\frac{1}{3}}+(1 2 1)_{-\frac{1}{2}} + (1 2 1)_{\frac{1}{2}}  
  + (1 1 1)_{0}  + (1 1 1)_{0}
\end{eqnarray}

The third family (\ref{3rdFamily})
\begin{eqnarray}
&&(3 \bar{3} 1 1) \rightarrow (3 2 1)_{\frac{1}{6}} + (3 1 1)_{-\frac{1}{3}}\\\nonumber
&&(1 3 \bar{3} 1) \rightarrow (1 2 1)_{-\frac{1}{2}} + (1 2 1)_{\frac{1}{2}} + (1 2 1)_{\frac{1}{2}} + (1 1 1)_{0} + (1 1 1)_{1} + (1 1 1)_{0}\\\nonumber
&&(1 1 3 \bar{3}) \rightarrow (1 1 2)_{0} + (1 1 2)_{ -1} + (1 1 2)_{0 } + (1 1 1)_{1} + (1 1 1)_{0 } + (1 1 1)_{1}\\\nonumber
&&(\bar{3} 1 1 3) \rightarrow (\bar{3} 1 2)_{\frac{1}{3}} + (\bar{3} 1 1)_{-\frac{2}{3}}
\end{eqnarray}
which we rearrange  in a more suggestive form
\begin{eqnarray}
&&  (3 2 1)_{\frac{1}{6}}   + (\bar{3} 1 1)_{-\frac{2}{3}} + (1 2 1)_{\frac{1}{2}} + (1 1 1)_{1}  \label{Qfam}  \\\nonumber
&&+(\bar{3} 1 2)_{\frac{1}{3}} +[(1 1 2)_{0}+ (1 1 2)_{0 } ]+  [ (1 1 2)_{ -1}+ (1 1 1)_{1}  + (1 1 1)_{1}]+ (1 1 1)_{0 } \\\nonumber
&&  + (3 1 1)_{-\frac{1}{3}}  + [(1 2 1)_{-\frac{1}{2}}  + (1 2 1)_{\frac{1}{2}}] + (1 1 1)_{0}  + (1 1 1)_{0}
\end{eqnarray}
The first line of Eq.(\ref{Qfam}) contains a SM family except that $b_R$ is missing. The second line contains some states in nontrivial $SU(2)_l$ irreps, some of which are in nontrivial $SU(2)_l$ irreps, and the last line contains the remaining states.

In order to complete the third SM family, we can either (i) break $SU(2)_l \rightarrow 0$ at a scale $M_{ssb}$, or (ii) arrange to have the gauge coupling of $SU(2)_l$  run to large values, where at some scale $\Lambda_{l}$  this group becomes confining. We expect the lower bounds on  $M_{ssb}$ and $\Lambda_{l}$ to be similar.

\subsection{Completing the third family via spontaneous symmetry breaking}

Let us  discuss the case of spontaneous symmetry breaking (i) first. We introduce a scalar $SU(2)_l $ doublet  $(1,1,2)$ who's VEV breaks $SU(2)_l$ completely so that $(\bar{3} 1 2)_{\frac{1}{3}} \rightarrow (\bar{3} 1 1)_{\frac{1}{3}}+(\bar{3} 1 1)_{\frac{1}{3}}$. One of these two irreps
can be identified with the $b_R$, hence  completing the third family in the first line of Eq.(\ref{Qfam}). The other we identify as the  $b'_R$, which  pairs with the $(3 1 1)_{-\frac{1}{3}} $ in the third line of Eq.(\ref{Qfam}). The chargeless $SU(2)_l$ doublet leptonic states in the second line of Eq.(\ref{Qfam}) also split into singlets, while the charge -1 doublet $SU(2)_l$ irrep splits so that they can pair with the charge +1 singlet leptons in that line. 
Writing Eq.(\ref{Qfam}) after the symmetry breaking, where we have  moved half the split $(\bar{3} 1 2)_{\frac{1}{3}} $ irrep into the first line and the other half into the third line gives
\begin{eqnarray}
&&  (3 2 1)_{\frac{1}{6}}  +(\bar{3} 1 1 )_{\frac{1}{3}}  + (\bar{3} 1 1)_{-\frac{2}{3}} + (1 2 1)_{\frac{1}{2}} + (1 1 1)_{1}  \label{Qfam2}  \\\nonumber
&& +[(1 1 1)_{0}+ (1 1 1)_{0 } +(1 1 1)_{0}+ (1 1 1)_{0 } ]+  [ (1 1 1)_{ -1}+  [ (1 1 1)_{ -1}+ (1 1 1)_{1}  + (1 1 1)_{1}]+ (1 1 1)_{0 } \\\nonumber
&& +(\bar{3} 1 1)_{\frac{1}{3}}  + (3 1 1)_{-\frac{1}{3}}  + [(1 2 1)_{-\frac{1}{2}}  + (1 2 1)_{\frac{1}{2}}] + (1 1 1)_{0}  + (1 1 1)_{0}
\end{eqnarray}
This has yielded a standard third family in the first line,    states with identical charges to the extra trinification family in the third line,  plus the new extra states of a quartification family in the second line.  It is the properties of the  $b$ quark that will interest us most.

Note that all three families have an extra $d'$ type quark in $(3 1 1)_{-\frac{1}{3}}+(\bar{3} 1 1)_{\frac{1}{3}}$, which is typical of all trinification or $E_6$ models. For the first two families they are in vectorlike representations, so these particles can acquire mass at a high scale, and we will not discuss them further. However, in the third family the $b'$ can not acquire a mass until $SU(2)_l$ is broken. Thus the third family $b'$ is phenomenologically more interesting.

\subsection{Completing the third family via a confining unbroken $SU(2)_l$}

Now let us discuss the third family for case (ii), a confining unbroken $SU(2)_l$.
All  necessary SM third family states  are in  $SU(2)_l$ singlets except for the  $(b_{R})$. A potential candidate  is in an $SU(2)_l$ doublet, the state $(\bar{3} 1 2)_{\frac{1}{3}}$ which contains part of the $b$ quark, but since it is an $SU(2)_\ell$ doublet, it must be confined before it can be identified with the $b_R$. This can be done by binding it to a $(112)_0$ scalar to generate a massless composite $$(\bar{3} 1 1)_{\frac{1}{3}}\sim (\bar{3} 1 2)_{\frac{1}{3}}\times (112)_0.$$
This leads to some interesting phenomenological consequences,  as we will discuss in the next section.
For this particle to be in a hadron we must confine it again via color $SU(3)_C$. Hence there is a double confinement process and a double hadronization, first (assuming $\Lambda_l > \Lambda_{QCD} $) via 
$SU(2)_\ell$ and then via $SU(3)_C$. (In standard quartification, none of the SM fermions need  leptonic confinement or leptonic hadronization.) 

Note that there are only four fermionic $SU(2)_\ell$ doublets, so the $SU(2)_\ell$ beta function indicates that it is a confining gauge theory as required, and if it starts off with a coupling $g_l>>g_3$ at the $SU(3)^4$ scale, then it is possible for $SU(2)_\ell$ to confine before  $SU(3)_C$. Likewise the fermionic spectrum is also consistent with $SU(3)_C$ being confining. Similar remarks apply to the $SU(2)_l$ doublet leptons which must be confined for large $g_l$. Hence their natural mass scale is around $\Lambda_l$.

\section{Phenomenological Implications and Summary}

We now discuss the phenomenology of our two models. 

\subsection{SSB Phenomenology}

For case (i), with spontaneous symmetry breaking of $SU(2)_l$, we find a phenomenology that is a straightforward extension of the SM: it contains the normal SM particle content   in the first two families plus their trinification extension. The third quartified family contains a third normal family, its  extended trinification content plus the remaining  extended quartification content composed of two $SU(2)_L$ singlet unit electric charged leptons and five Weyl neutrinos some of which can be pair up after SSB. While this model is potentially interesting, it is not particularly novel and further analysis and predictions would proceed along standard lines for the SM plus additional particle content. What is most interesting is the case when $SU(2)_l$ becomes confining as we now discuss.

\subsection{$SU(2)_l$ Confinement Phenomenology}

Our case (ii) model contains 
the fundamental charge -$\frac{1}{3}$ quark in a doublet of the strong $SU(2)_l$ group.  If this particle binds with a charge neutral scalar $SU(2)_l$ doublet, as assumed herein for definiteness, then a composite particle in an $SU(2)_l$ singlet can form   with standard $b_R$ quark quantum numbers. This composite object should behave like a point particle up to some energy scale.  There are high-energy scatterings that set a lower bound on this scale such as $e^+ e^- \rightarrow b\bar{b}$ which have shown $b_R$ to behave like point particles up to $M_Z$, but above that scale anomalies could appear directly, or below $M_Z$ indirectly.
  
Since $SU(2)_l$ is confining, the model is expected to have both $SU(2)_l$ mesons and baryons, where as usual, the mesons are composed of an $SU(2)_l$ doublet and a conjugate doublet, while  the baryons are composed of two $SU(2)_l$
doublets (only two quarks for Muster Mark this time!), but since the doublet is a pseudoreal irrep, the distinction between mesons and baryons is blurred. 
There can also be $SU(2)_l$ glueballs in the model. All these states should be near the $SU(2)_l$ confinement scale.

The model's phenomenology divides according to the energy transfer which probes the composite nature of the  $b_R$. We will consider the two cases where the probe energy compared to the binding energy of the composite $b_R$ is small or large. (The third case of a probe energy comparable in magnitude to the binding energy is a more subtle proposition, and not considered here.)
If the probe energy is small, then (unknown) form factors govern the process.  These form factors may present themselves as BSM physics. (In an alternative approach \cite{Kayser:1996hu}, the CP properties of the $B$ states are emphasized, and it is unclear to these authors how a composite $b_R$-quark would enhance effects.) 
If the probe energy exceeds the unknown binding energy, then the composite $b_R$-quark is resolved into its constituents, 
the fundamental precursor to the $b$-quark and the scalar. The binding energy would naturally be the order of the the confinement scale $\Lambda_l$, possibly as low as a few GeV as governed by upsilon phenomenology, but probably of order of 100 GeV, much like the conventional Higgs doublet.
What is different from the usual SM case is that there is no light left-over fundamental $SU(2)_{l}$ Higgs in this case. 

The $b_{L}^{c}$ state is the charge conjugate of the right handed component of the $b$ quark.  Since any free particle must be a singlet of $SU(2)_l$ below its confinement scale, then in any $b$-onium state the $b_{L}^{c}$ must be bound to a state with opposite lepton color charge below that scale.  The standard $\Upsilon$ is taken to be $b\bar{b}$, but within the present model it could be formed from the binding of two pairs of bound objects: the $b_R$ now will form a composite from binding with a (112) scalar, and then together they combine with the anti-$b_{L}^{c}$ and anti-(112) scalar (along with the left handed piece and its anti-left pair) to form a lepton colorless particle. The additional structure could then be probed with energies exceeding binding energy of the $b$+scalar composite.

Recently, 
a number of anomalous results in $B$ decays (and so by implication, $b$ decays at a more fundamental level) have been reported. None of these observations are
convincing on their own, but taken together as a whole it does appear
that they imply new physics
beyond the standard model (BSM) is required to explain these new results. These results indicate a violation of Lepton flavor universality (LFU) of the coupling of electroweak gauge bosons to all 
three families of leptons, a consequence of the Standard Model (SM) \cite{Graverini:2018riw}.
Global analyses and reviews of these $B$ anomalies include \cite{Dutta:2002ah,Hurth:2013ssa,Dumont:2016xpj,Buttazzo:2017ixm,Angelescu:2018tyl,Kumar:2018kmr,Cornella:2019hct,Arbey:2019duh,Descotes-Genon:2015uva,London:2019nlu}, and references therein. One possible explanation for the $B$ anomalies is through partial compositeness, which can provide violation of LFU (see, for example \cite{Niehoff:2015bfa,Sannino:2017utc,Stangl:2019esx}). These models typically have a composite Higgs which leads to degrees of compositeness for mutliple quarks, as well as mediators and muons. In contrast, the flipped quartification model in the present work does provide compositeness for only the $b_R$-quark, but through a new confining gauge group.

While we are not ready to provide an explanation of the anomalies here, we have presented a 
model that has the potential of eventually doing so, i.e., our Flipped Quartification model,    is able to single out the $b_R$ quark as the only fermion among the three families that differs from its designation in 
the SM. There are two viable phenomenological options in the model, the $b_R$ must either (i) couple to a new low scale gauge group $SU(2)_l$ that breaks in a way so that the $b$ acquires its usual SM quantum numbers and remains a fundamental particle, or (ii) the new $SU(2)_l$ gauge group becomes confining and the $b_R$ becomes a composite upon binding with a scalar.

While the model we have presented focuses
on $B$ physics, other models in this class can be used to 
single out one or more right handed charge $-\frac{1}{3}$
quarks. Then right handed quarks  are made to fall into flipped quartification
families, while the remaining right handed charge $-\frac{1}{3}$
quarks remain in trinification families. 
Future work potentially leads to a whole class of models similar
to   Flipped Quartification   where one or more fermions are singled out 
to differ from other normal family members, 
hence providing a rich and interesting BSM phenomenology.

\section*{Acknowledgments}
We thank Gudrun Hiller for useful discussions about partial compositeness and flavor anomalies.
This work was supported in part by the US Department of Energy under Grants
DE-FG05-85ER40226 (JBD and TWK), DE-SC-0019235 (TWK), DE-SC-001198(TJW), and  DE-FG03-91ER40833 (HP). TWK and HP thank the Aspen Center for Physics for hospitality where this research was initiated some time ago. JBD acknowledges support from the National Science Foundation under Grant No. NSF PHY182080
 
\bibliography{main}

\end{document}